# Hydride growth mechanism in Zircaloy-4: investigation of the partitioning of alloying elements

Isabelle Mouton[1]*, Yanhong Chang[1], Poulami Chakraborty[1], Siyang Wang[2], Leigh T. Stephenson[1], T. Ben Britton[2], Baptiste Gault[1,2,*]

[1.] Max-Planck-Institut für Eisenforschung, Max-Planck-Straße 1, 40237 Düsseldorf, Germany.
[2.] Department of Materials, Royal School of Mines, Imperial College London, London, SW7 2AZ, UK
* Corresponding author: isabelle.mouton42@gmail.com, b.gault@mpie.de

## Abstract

The long-term safety of water-based nuclear reactors relies in part on the reliability of zirconium-based nuclear fuel. Yet the progressive ingress of hydrogen during service makes zirconium alloys subject to delayed hydride cracking. Here, we use a combination of electron back-scattered diffraction and atom probe tomography to investigate specific microstructural features from the as-received sample and in the blocky-α microstructure, before and after electrochemical charging with hydrogen/deuterium followed by a low temperature heat treatment at 400 °C for 5 hours followed by furnace cooling at a rate of 0.5 °C/min. Specimens for atom probe were prepared at cryogenic temperature to avoid the formation of spurious hydrides. We report on the compositional evolution of grains and grain boundaries over the course of the sample's thermal history, as well as the ways the growth of the hydrides modifies locally the composition and the structure of the alloy. We observe a significant amount of deuterium left in the matrix, even after the slow cooling and growth of the hydrides. Stacking faults form ahead of the growth front and the segregation of Sn at the hydride/matrix interface and on these faults. We propose that this segregation may facilitate further growth of the hydride. Our systematic investigation enables us discuss how the solute distribution affects the evolution of the alloy's properties during its service lifetime.

## 1. Introduction

The primary application of zirconium-based alloys is in nuclear fuel cladding for water-based reactors [1]. This is because Zr has a low neutron absorption cross section, and its alloys exhibit good mechanical properties and a high resistance to corrosion. Zirconium has indeed a high oxidation potential and a layer of $ZrO_2$ readily forms upon contact with air and grows over time in the water coolant, which releases hydrogen. However, the release of hydrogen associated to this oxidation process can lead to a progressive ingress of hydrogen in the underlying alloy [2]. Additional hydrogen can also be released from the nuclear fuel. A consequence of this increase in hydrogen concentration is that these alloys are prone to hydrogen embrittlement (HE) via delayed hydride cracking (DHC) or hydrogen-induced delayed cracking (HIDC) [3–5]. Indeed, the solid solubility of hydrogen in α-Zr is lower than 10 wt. ppm at ambient pressure and room temperature [6], and so the driving force for the precipitation of zirconium hydride is very high. Four zirconium hydride phases have been reported in the literature, with different crystal structure and range of stoichiometries $ZrH_x$ [7–9], such as hexagonal close packed (HCP) ζ (x=0.25-0.5), face-centred tetragonal (FCT) γ (x=1), face-centred cubic (FCC) δ (x≈1.5-1.65) and FCT ε (x≈1.75-2).

Significant uncertainty also exists in the thermodynamics of the Zr-H system and variation in the proposed phase diagrams is evident, particularly below the eutectic temperature of 550 ˚C [2,10,11]. These hydrides are known to be mechanically brittle [12]. There are conflicting reports regarding whether cracks initiate within the hydride and/or the hydride/α-Zr interface [13,14]. It is hence not unexpected that the volume fraction and size of the hydrides are amongst the crucial parameters in controlling the embrittlement process [4,15,16]. Other parameters such as the orientation of the hydride with respect to the load axis are also known to be of importance [12].

At room temperature, Zircaloy-4 comprises a hexagonal close-packed (HCP) α phase matrix, with typically a 0.5 % volume fraction of secondary phase particles. Recent efforts to control the grain size and morphology have resulted in the possibility to generate a coarse-grained (grain size typically > 200 µm) microstructure, referred to as 'blocky-α', from the as-received microstructure with a grain size of approx. 11 µm [17]. In part this was motivated by trying to limit the grain boundary area and hence assist in understanding the nucleation of hydrides. Two types of intergranular hydrides were reported, with a growth either parallel (a-type) or perpendicular (b-type) to the grain boundary [18], in addition to intragranular hydrides. Hydrides were also reported at the boundary of deformation twin boundaries formed during processing at room temperature[18].

While the presence of hydrides in Zr alloys has been largely observed and commented in the literature [19–21], however aspects of the growth mechanism are still elusive, partly because experimental observation at the near-atomic scale is scarce. Breen et al. recently studied specifically the growth front of hydrides formed in Zircaloy-4 (Zr-1.5%Sn-0.2%Fe-0.1%Cr wt%), electrochemically charged with hydrogen and thermally treated to form hydrides [22] (homogenisation and subsequent solid phase precipitation). They showed that an intermediate interface between the hydride and the matrix can has its own atomic structure and interfacial chemistry, which was also expected to play a critical role in the nucleation and growth processes. Another important aspect of this preliminary work is the interaction between the growth of the hydrides and the alloying elements. Fe and Cr were previously reported to form secondary-phase particles, i.e. Laves phase, or to segregate to grain boundaries (GBs) [5,23–25], each with a distinct influence on the corrosion and mechanical properties.

To further our understanding of the nucleation and growth mechanism of zirconium hydrides formed through the solid state Zr+[H] → Zr + hydride transus in Zircaloy-4, atom probe tomography (APT) is employed in the present work. The chemical sensitivity and high spatial resolution of APT makes the technique well suited to detect segregation of alloying additions to interfaces for example and to complement scanning electron microscopy and electron backscattered diffraction (EBSD), which were used primarily for target preparation of GBs with known misorientation, later be correlated with their composition. We examined selected grain boundaries in three different sample: (1) as-received, (2) blocky-α [17], and a (3) H/D charged blocky-α Zircaloy-4. Of particular interest is the distribution of Fe and Sn at grain boundaries in fine and blocky-α Zircaloy-4, which appears unchanged in the two samples. Fe is seen to form small clusters in the α-Zr matrix. After hydriding, the characteristics of Fe distribution in the grain interior and at the grain boundary remain unaffected. Sn, however, is repelled from the growing hydrides. In the hydrided sample, we first targeted a twin boundary, which was reported to be a prime nucleation site for hydrides in this alloy [26]. We also focused on the growth front of the b-type intergranular hydride, which shows interesting deformation features in the form of stacking-faults that appear segregated with Sn. Our results shed light onto solute-hydride interactions that help control the growth of hydride, which could help the design of alloys with longer service lifetime and towards higher fuel burnup.

## 2. Materials and methods

### 2.1. Materials

The starting alloy studied is a commercial Zircaloy-4 from a rolled and recrystallized plate with a composition of Zr-1.5Sn-0.2Fe-0.1Cr wt%. It is supplied with a recrystallised grain structure, where average grain size is ~11 µm. The material was then heat treated at 800˚C for two weeks; in order to obtain large blocky-α grains of >200 µm in average size, as described in Tong and Britton [17]. Samples were electrochemically charged with hydrogen or deuterium using galvanostatic charging at a current density of 2 kA/m$^2$, using a solution of 1.5 wt. % H/D$_2$SO$_4$ in H/D$_2$O at 65 ˚C for 24 hours. Charging conditions were selected based on prior work [22].

An approximately 20µm thick hydride layer was formed at the surface [22]. The H/D was then homogenised inside the bulk via annealing at 400 ˚C for 5 hours, followed by furnace cooling at a rate of 0.5 ˚C/min to promote the formation of δ-hydrides. This process tends to results in the formation of hydrides at grain boundaries and within the blocky-α grains. The samples, after H or D charging and diffusion heat treatment, were ground with 1200 (1 min), 2000 (3 min) and 4000 grit (5 min) (FEPA designation) SiC paper, using water as lubricant, and then polished mechanically on a Struers MD-Chem polishing cloth with a solution of OPS:H$_2$O$_2$:H$_2$O=5:1:6 (vol.) for 1.5 h. Heavy water was not used for the lubricant in part due expense, but mostly because we targeted precipitated hydrides or deuterides. Electropolishing was then performed with HClO$_4$:methanol=1:9 (vol.) with 30 V applied voltage at -40°C for 90s, and the conditions here mean that this step can only remove and not add surface hydrides.

Analysis was performed on samples at different stages during preparation, as detailed in Table 1.

Table 1:

| Specimen | Thermal treatment | Uncharged/Charged | Site specific |
|---|---|---|---|
| A | as-received fine grain | - | Grain boundary |
| B,C | Annealed blocky-α large grain | - | Grain boundary |
| D* | Annealed blocky-α large grain | Deuterium charged | - |
| E, F,G | Annealed blocky-α large grain | Hydrogen charged | Twin boundary hydride |
| H | Annealed blocky-α large grain | Deuterium charged | Intergranular hydride |
| I, J | Annealed blocky-α large grain | Hydrogen charged | Intergranular hydride |

\* Milling performed at cryogenic temperature

### 2.2. EBSD characterization

Electron backscattered diffraction (EBSD) was performed with a FEI Quanta 650 scanning electron microscope (SEM) equipped with a field emission gun operated at 20 kV and a probe current around 10 nA, using a Bruker eFlashHR (v2) detector and the Esprit 2.1 software. Bruker ESPRIT 2.1 software was used for the collection and analysis of data. Patterns were binned to 320×240 pixels in order to decrease the exposure time therefore to avoid excessive sample drifting, enabling a 22ms exposure time per pattern.

### 2.3. Atom probe tomography specimen preparation

APT specimens were prepared using a FEI Helios dual-beam xenon plasma focused ion beam (PFIB), following the procedure introduced by Thompson et al. [27]. Most specimens were prepared at ambient temperature at an acceleration voltage of 30 kV and milling currents between 0.46 nA~24 pA and a final cleaning at 2 kV and 24 pA. Hydride formation during FIB milling has been reported in Ti- based alloys, and the FIB-milling is also known to potentially affect the nature of the hydride [28,29]. Here, we also face a problem of hydride formation in Zr-based alloys.

APT specimen prepared conventionally by PFIB at ambient temperature on uncharged blocky-α sample can either:

- Have no influence on the microstructure and the HCP crystallography but increase the composition of H in the matrix, as observed in Figure 1 a;

- Induce the formation of a small hydride phase, often positioned at the edge of the APT specimen;

- Completely transform the initial HCP phase into a hydride phase. The corresponding APT detector hit map, or detector density histogram, as shown in Figure 1 b, exhibiting the typical symmetries of a FCC-like phase, which can be associated with one of the hydride phases. The spatial resolution of APT does not allow us to distinguish between FCC and FCT.

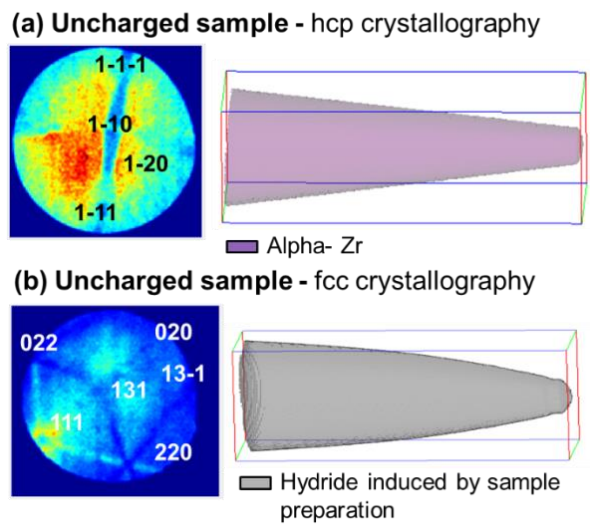

**Fig. 1.** APT desorption map, reconstruction and mass spectrum the same uncharged sample in the α -Zr phase (a) without and (b) with influence of FIB sample preparation.

Fig. 2 depicts examples of possible alterations of the hydride or of the distribution of solute elements. The APT data in Figure 2 was obtained from an intergranular deuteride previously located by a combination of SEM and EBSD. The APT reconstruction, Figure 2 a, shows three different hydride phases with different crystallographic orientations, as revealed by the detector hit maps in Figure 2 b, c and d, as well as a small grain of the α-phase. The hydrides we are interested in studying, based upon the sample treatment, should be deuterium rich and therefore an important observation here is that in comparison with other deuterated phase analysed by APT [30,31], in this volume, there is a lack of deuterium and instead they are hydrogen rich, indicating that they were introduced during sample preparation. Prior work indicates that FIB-based specimen preparation at ambient temperature affects the nature of the hydride [32]. Additionally, to the characteristic atomic density and compositional variation in the FIB-induced hydrides, there is an absence of Sn enrichment near and at the α/hydride interface, which was observed by Breen et al. [30]. This confirms that the lack of Sn is important in the deuteride precipitation during the thermal treatment, and that any hydrides or deuterides without this signature are likely artefacts of sample preparation.

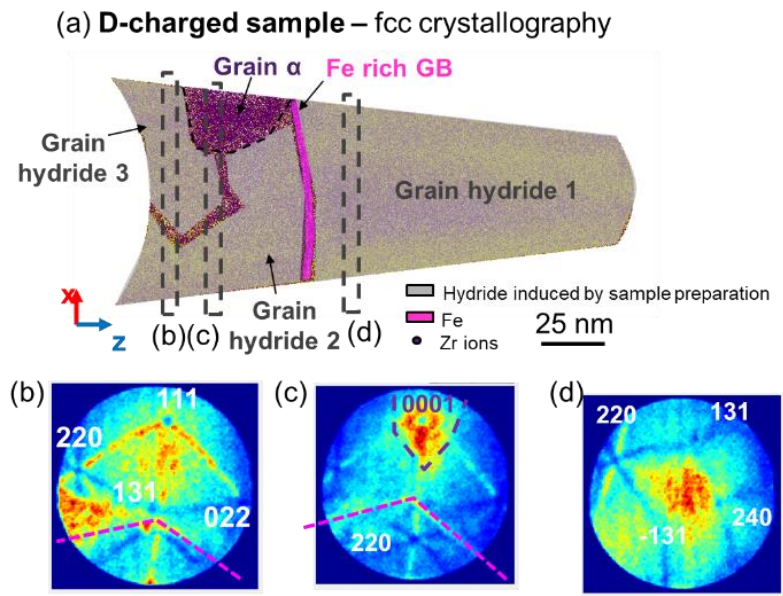

**Fig. 2.** Slice from APT reconstruction performed on D charged sample. The nature of the deuderide potentially present were affect with the FIB sample preparation.

Chang et al. [32] reported a protocol to prevent hydride phase formation using cryogenic preparation. Although the lift-out part is performed at ambient temperature, the sharpening and final polishing of the APT specimens is performed under cryogenic conditions on the setup described in ref. [33]. Several APT experiments were also performed in the present manuscript with specimen prepared at cryogenics temperature, in this case the APT detector hit maps always shows the HCP crystallography of the initial α -Zr phase with lower amounts of H within the α-Zr matrix. Since we derived a way to distinguish between the hydrides that are FIB-induced and those from the sample, we choose to perform our experiments mainly with specimens prepared at ambient temperature. APT experiments with obvious or suspected FIB-induced hydrides were discarded. If small hydrides, or hydrides at the edge of the dataset were identified but do not interfere with the microstructural features of interest, they are represented in grey otherwise hydride from H/D charging are represented in yellow. Finally, as the preparation introduces H in solution, the composition in H will be not discussed in detail herein.

### 2.4. Atom probe tomography characterization

All the specimens were analysed on a CAMECA LEAP 5000 XR, at a base temperature of 50K. The instrument was operated in laser pulsing mode. Quantification of hydrogen by APT is not trivial. There is a growing body of work in the literature showcasing the possibility to measure the content of hydrogen contained in hydride phases as well as solute H in a matrix[29–31,34]. Results are made complicated by the residual H from within the APT chamber which ends up being detected as part of the analysis, and by the possibility that neutral hydrogen or dihydrogen detaches from a field evaporated molecular ion during the ion's flight between the specimen and the single-particle position-sensitive detector used in the atom probe. For the characterization of Zr-based alloys, we discussed these issues in details in Mouton et al. [31,35], where we also introduced a protocol to correct the composition profiles to account for some of these issues. These methods were also used here, and the interested readers are encouraged to refer to this previous study. Please note that since the results appear comparable, we do not directly discriminate between hydrides and deuterides, and use the term interchangeably, except for the section of the discussion regarding the matrix composition in 4.1.

## 3. Results

### 3.1. As-received Zircaloy-4

The EBSD maps in Fig. 3 a and b present an overview of the crystallographic orientation of the grains in the as-received Zircaloy-4. The grain size in the as-received sample is approx. 11 µm, which exhibits a typical split basal texture. APT specimen (*A*) was prepared in order to include a high angle grain boundary (Fig 3.c) previously chosen from the EBSD map (Figure 3 b). The grain boundary in the APT reconstructed volume in Figure 3 d is highlighted by an iso-composition surface delineating the regions containing more than 1.5 at% of Fe. Figure 3.e plots the composition profile calculated perpendicularly to the grain boundary. This profile brings out both composition in Fe atoms up to 2 at.% and also a slight enrichment in Sn (up to approx. 1.4 at.%) at the grain boundary compared to the surrounding matrix that contains only approx. 0.2 at.%  and 0.9 at.% of Fe and Sn respectively.

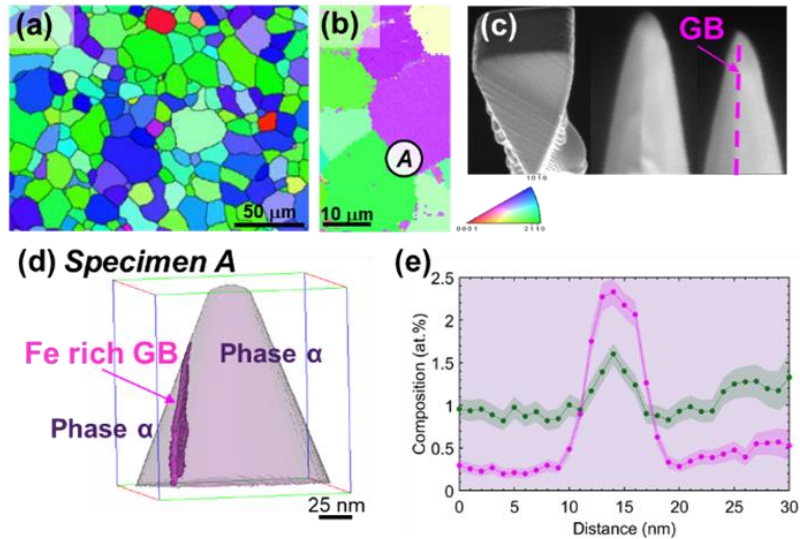

**Fig. 3.** (a) EBSD inverse pole figure (IPF) map of **As-received Zircaloy-4, coloured with respect to the out of plane direction** (b) EBSD IPF map of APT specimen, **coloured with respect to the out of plane direction** (*A*) localisation on grain boundary of interest (c) APT specimen preparation in order to include high angle grain boundary (d) APT reconstruction of the *specimen A.* (e, f) Composition profile performed perpendicularly to the grain boundary.

### 3.2. Annealed Zircaloy-4 blocky-α

Fig 4.a shows the EBSD map of the Zircaloy-4 after heat treated at 800°C for two weeks, leading to a microstructure referred to as blocky-α with an average grain size of over 200μm. Several APT specimens were prepared along a high angle grain boundary (Figure 4.b). APT reconstructions of the datasets obtained from the analysis of *specimen B* and *specimen C* are shown in Figure 4 c and d, respectively. The reconstruction of *specimen C* contains a small hydride induced by specimen preparation shown in grey on the Figure 4 d. Composition profiles across the grain boundary in both datasets evidence an Fe enrichment to the grain boundary, with compositions ranging from 0.8 at.% to 1.8 at.% Fe, i.e. lower than the grain boundary in the as-received Zircaloy-4 (Fig 3). The Sn composition is approx. 1.8 at.% in both datasets. The Sn matrix composition is the same as in the as-received alloy, but the Fe present in solid solution is reduced from approx. 0.25 at% to almost 0 (comparable to the level of background).

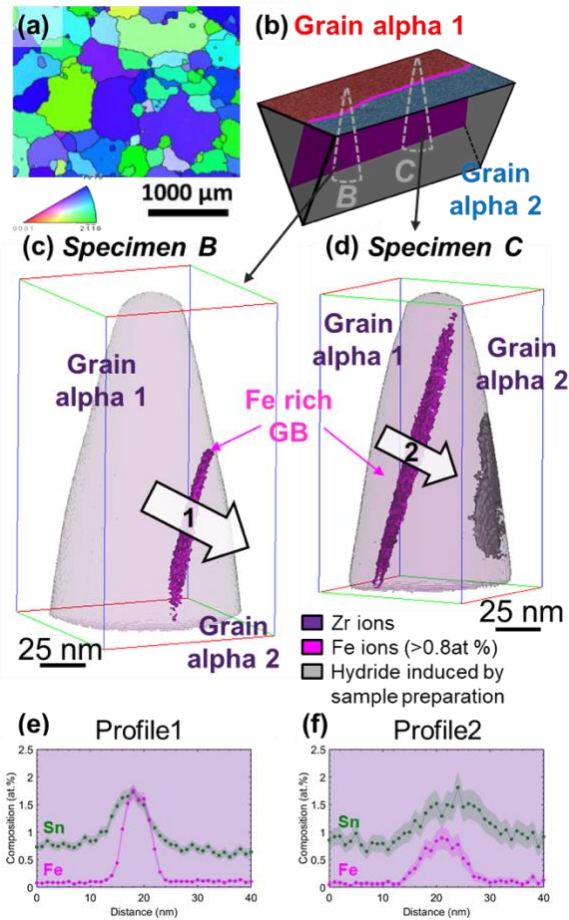

**Fig. 4.** (a) EBSD IPF map of **Zircaloy-4 blocky-α, coloured with respect to the out of plane direction** (b) APT specimen preparation along the same grain boundary (c and d) APT reconstruction of the *specimen B* and *C*. (e and f) Composition profiles performed perpendicularly to the grain boundary.

Another APT specimen, labelled *D*, was prepared from the centre of a blocky-α grain, to evaluate the Fe distribution far from a grain boundary. The distribution of Fe is inhomogeneous throughout the reconstruction as shown in Figure 5 a. A composition profile in the form of a proximity histogram (i.e. proxigram) as a function of the distance to a 2 at.% isocomposition surface was performed on an individual cluster and is plotted in Figure 5 b. The composition at the core of this cluster reaches approx. 5 at% Fe. This particular dataset was obtained from a specimen prepared at cryogenic temperature. Similar Fe rich clusters were also observed away from grain boundaries in H/D charged samples.

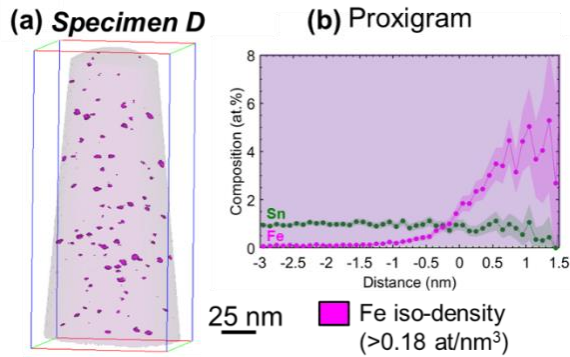

**Fig. 5.** (a) APT reconstruction of the *specimen D* from **Zircaloy-4 blocky-α** sample showing Fe rich clusters. (b) ) Proxigram performed on one cluster.

### 3.3. Microstructure in H/D charged blocky-α Zircaloy-4

It has been recently reported that H/D charged Zircaloy-4 sample shows, two types of (>µm sized) intergranular hydrides, described in two distinct type if the hydride is along (a-type) or perpendicular (b-type) to the grain boundary, that appear to be different to those growing at twin boundaries (for more details see ref [18]). At the length scale seen in the SEM or during FIB preparation, please note that what we tend to refer to as a hydride is often actually an assembly of smaller hydrides, i.e. a hydride packet. To prepare these APT specimens, we only targeted a specific hydride packet. Selecting a specific hydride grain within a packet is impossible since the typical scale of the microstructure inside these packets is smaller than what can be resolved during the final stages of specimen preparation. We selected one T1 twin boundary and another random high-angle grain boundary.

#### 3.3.1. Twin boundary hydride

Three specimens were prepared along the twin boundary hydride packet, as identified by EBSD, Figure 6 a. The *specimen E* shown in Figure 6 b is a (nearly) pure hydride needle that contains some Fe-rich clusters. A proxigram of one Fe-rich cluster, in Figure 6 c, reveals a composition at the core of more than 4 at% Fe, similar to those observed in the centre of the grain in Figure 5. The data obtained from *specimen F* contains two individual hydrided grains as well as an interface between the hydride and α-Zr (Figure 6 d). The composition profile calculated along arrow (1) in Figure 6 d which goes through the hydride-hydride interface and the hydride/α-Zr (Figure 6 e) shows that Sn segregates at both interfaces, with the segregation at the interface with the metallic matrix similar to previous reports [30,31]. In contrast, Fe does not segregate at either of these interfaces.

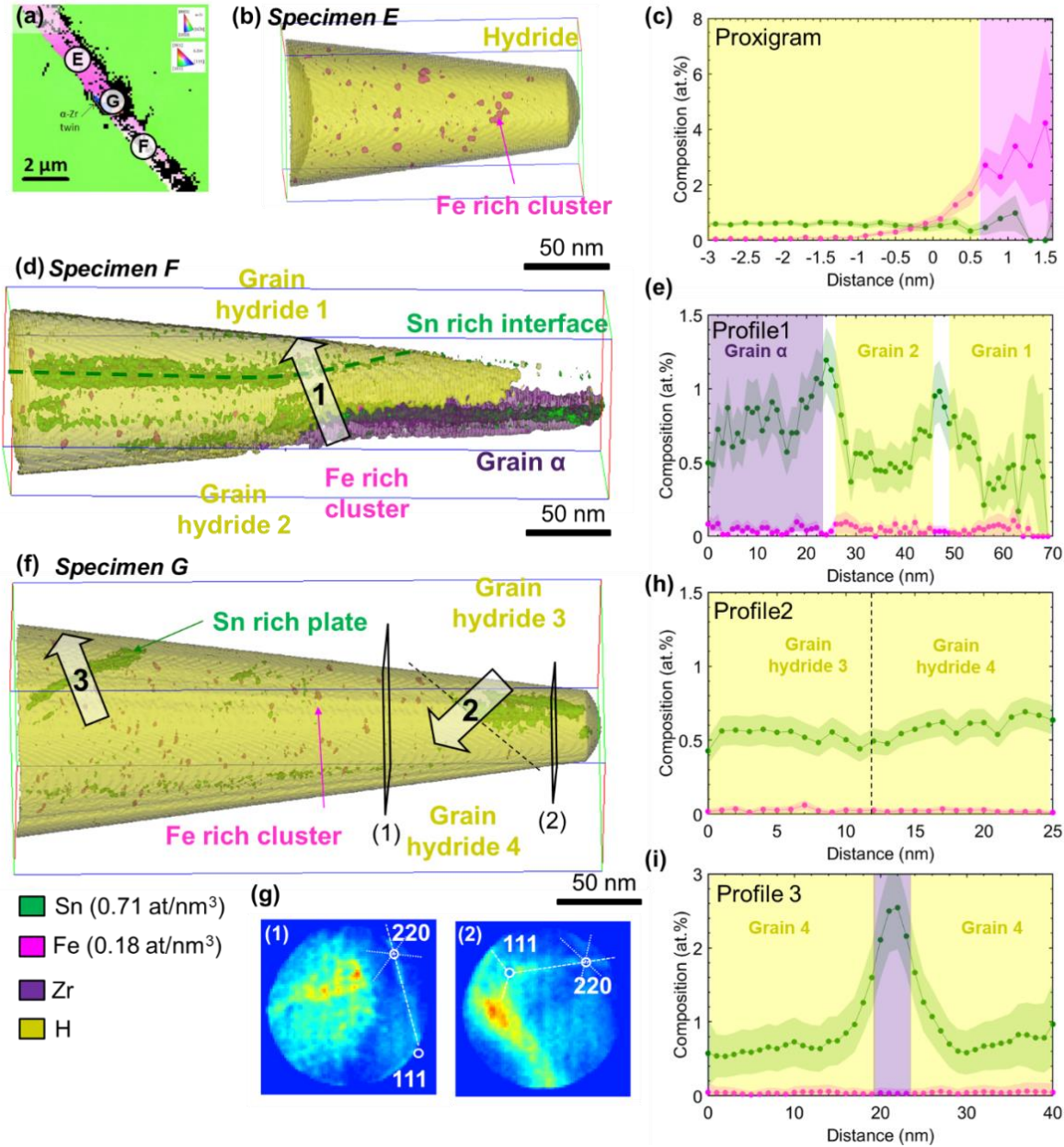

**Fig. 6.** (a) EBSD IPF map, coloured with respect to the out of plane direction, of a **twin boundary hydride in H/D charged blocky-α Zircaloy-4** (b,d and f) APT reconstruction of the *specimen E, F* and *G* respectively, position are indicated in (a). (c) Proxigram performed on one cluster of the *specimen E*. (e, f and i) Composition profile taken along the arrow indicated in (c) and (d).

S*pecimen G* was extracted close to where we expected the α-Zr twin boundary was located. The APT reconstruction from *specimen G* in displayed in Figure 6 f. Like in previous datasets, small Fe-rich clusters are present. Two other features are interesting. First, the detector hit maps (Figure 6 g) reveal a change in orientation, yet no compositional change is observed between grain 3 & 4 (Figure 6 h). We interpret this as two hydrides that have grown on either side of the twin boundary. The lack of segregation would be related to the lack of segregation at the twin boundary prior to the hydriding of the sample. Twins are low-energy boundaries and often the location of very low to moderate segregations [36], especially since these likely formed during the final stage of the preparation of the samples before hydriding, so with limited opportunities for activating solute diffusion[17]. In addition, a Sn-rich planar feature is also revealed, and the composition profile in figure 6 i calculated along arrow (2) in Figure 6 f, shows an enrichment up to 2.5 at. % Sn. There is no evidence from the detector hit map that this enrichment corresponds to yet another grain boundary. This enrichment could correspond to a stacking fault, as will be discussed in the next section.

### 3.3.2. Intergranular hydride (type-a)

Figure 7 summarises the APT analysis of *specimens H*, which was extracted at a random high-angle grain boundary bearing an a-type intergranular hydride. Two hydrides have grown on either side of the grain boundary, which is enriched in Fe, as highlighted by the pink iso-composition surface delineating regions containing more than 0.8 at% Fe in Figure 7 a. The analysis also contains a small grain of α-Zr that straddles on both sides of the grain boundary. Figure 7 b is a top-view from a slice within the APT atom map showing the location of the GB in between the two hydride grains, and the locations from which we calculated two composition profiles between hydride grain 1 and hydride grain 2 (arrow (1)), and between α-Zr and hydride grain 2 (arrow (2)). The composition profile in Figure 7 c shows a noticeable enrichment in Fe at the grain boundary, with a composition reaching approx. 0.75 at.% Fe, along with a slight enrichment in Sn up to approx. 0.75 %at from a base level of 0.55 at%. At the GB, the hydrogen composition is also a few at% lower than in the bulk of the hydride. The composition profile in Figure 7 d, although noisy due to the smaller sampled volume, also shows Fe segregation along with a slight depletion in Sn compared to within α-Zr, but at a similar level to that in between the two hydride grains. The composition in Sn in this small α-Zr grain is higher than in the hydride yet lower than in the bulk (Figure 5).

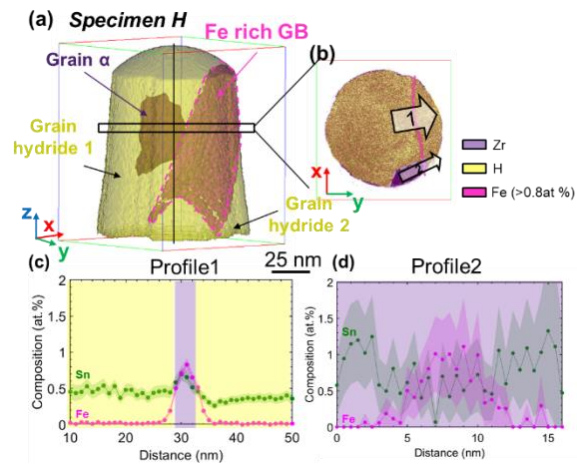

**Fig. 7.** (a) APT reconstruction of the *specimen H* including intergranular deuteride along a grain boundary from **in H/D charged blocky-α Zircaloy-4** sample. (b) Slice along xy from the APT reconstruction. (c and d) Composition profile performed perpendicularly to the grain boundary.

### 3.3.3. Intergranular hydride (type-b)

*Specimens I* and *J* were prepared from a b-type intergranular hydride, as shown in Figure 8 a. The APT reconstruction from *Specimen I* is shown in Figure 8 b. The reconstructed volume contains three distinct regions: a single, interconnected hydride phase in yellow and in between two regions containing α-Zr in purple. The two α-Zr grains have different crystallographic orientations, evidenced by a change in the pole pattern in the detector hit map. The Sn distribution is highlighted by a set of green iso-composition surfaces (>1.8 at.%), and this evidences a number of planar Sn-segregated thin plates in α-Zr. Fe rich cluster are also present but not shown here for clarity. The numbered arrows correspond to the regions from which composition profiles were calculated.

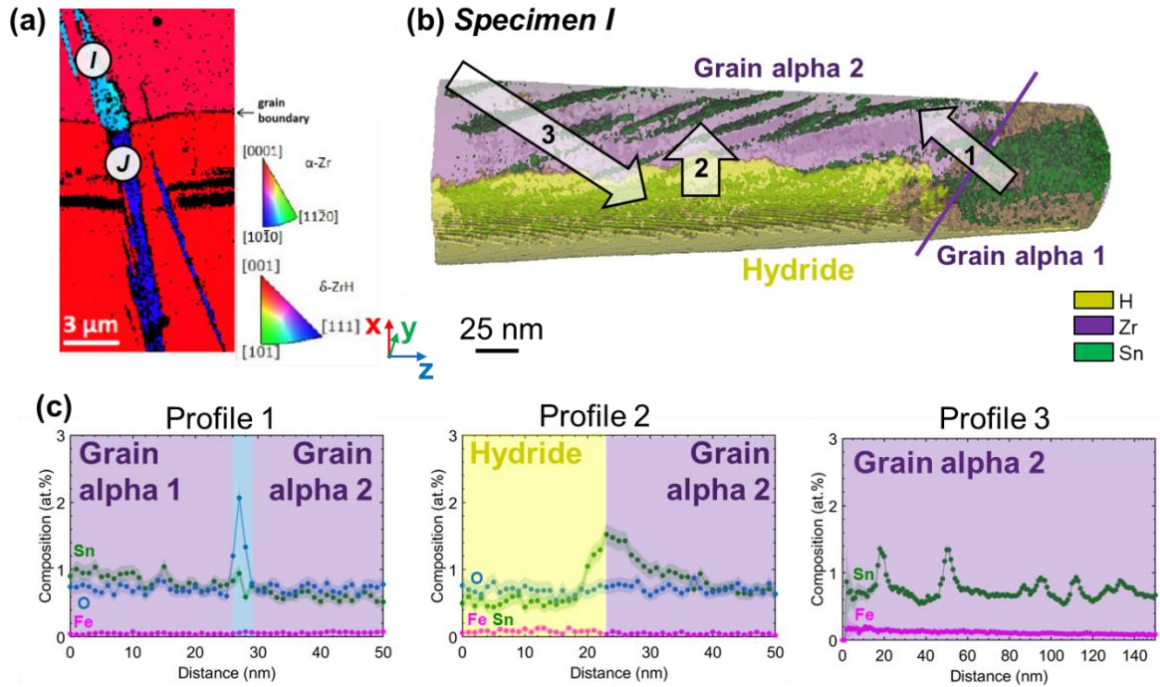

**Fig. 8.** (a) EBSD IPF map, presented with respect to the sample normal, of an **Intergranular hydride (type-b) in H/D charged blocky-α Zircaloy-4** (b) APT reconstruction of the *specimen I*, position are indicated in (a). (c) Composition profile taken along the arrow indicated in (b).

Composition profile 1 is through the boundary between α-Zr-grain 1 and α-Zr-grain 2, and shows no segregation of Fe or Sn. However, we observe an enrichment reaching up to approx. 2 at. % O. The oxygen composition had so far not been plotted in the other profiles as it showed no specific variation across phase and grain boundaries. The O reported here is mostly detected in the form of ZrO molecular ions, which were not detected in the other datasets. Profiles 2 is obtained at the interface between the hydrides and the α-Zr grains, i.e. the hydride growth front, and shows the segregation of Sn previously observed and reported [30]. Composition profiles 3 is calculated perpendicularly to the sets of Sn-segregated planar features. These are found in both α-Zr and are interpreted as stacking faults decorated by Sn. It has long been reported that the stress associated to the mismatch in the lattice parameters between the hydride and the host metallic matrix leads to plastic deformation [37,38]. These stacking faults appear parallel to each other in each of the grains but with different orientation between α-Zr-grain 1 and α-Zr-grain 2, as would be expected from stacking faults lying on a specific set of planes from within the α-Zr matrix. The profiles through these faults shows a segregation reaching 1.5 at. % Sn.

Interestingly, the matrix in between stacking fault appears depleted in Sn compare to the α-Zr composition reported in e.g. Figure 4 and 5.

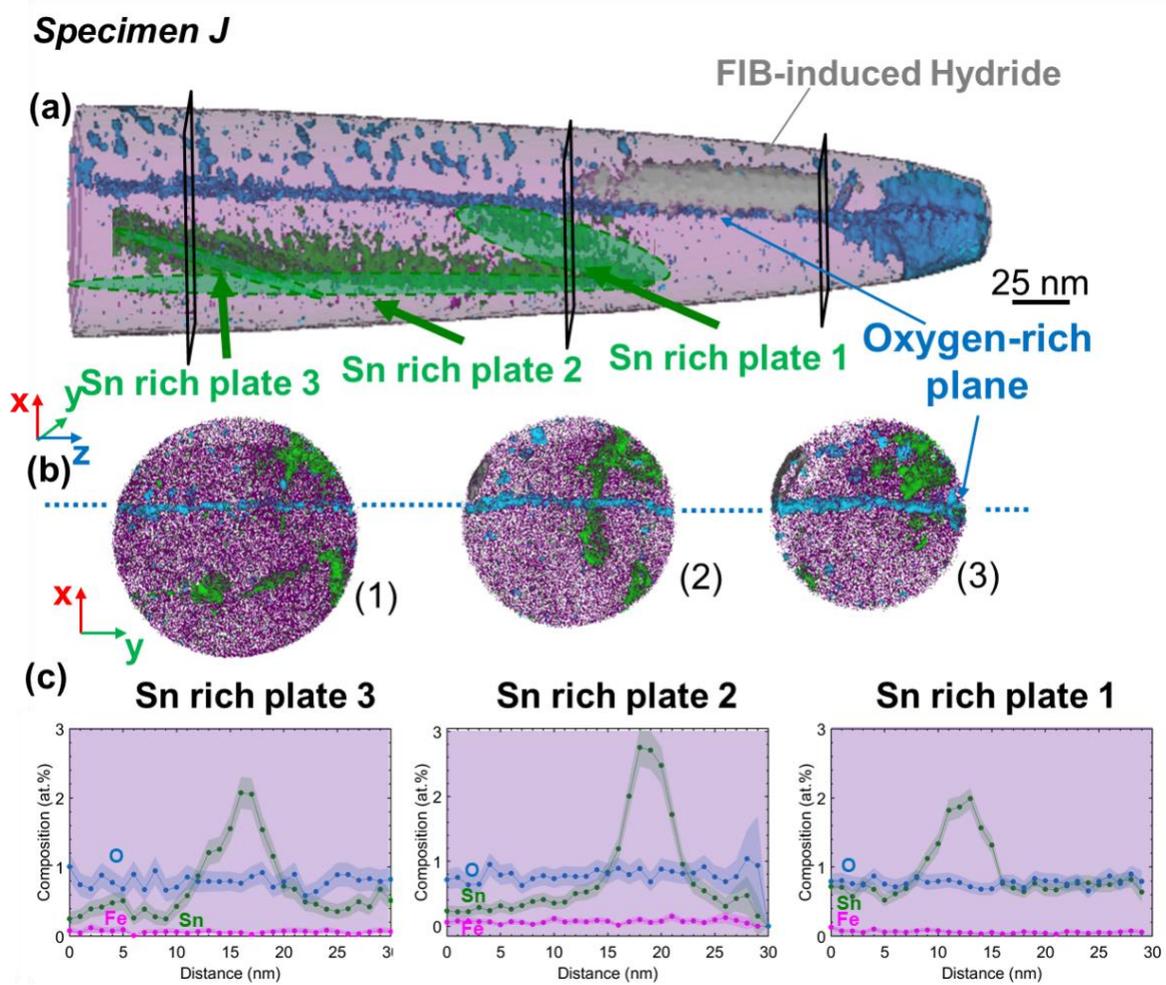

**Fig. 9.** (a and b) APT reconstruction of the *specimen J*, position are indicated in (a) in an **Intergranular hydride (type-b) in H/D charged blocky-α Zircaloy-4** (c) Slice along xy from the APT reconstruction at different z position indicated in (a). (d) Composition profiles taken perpendicularly to the Sn-rich plates.

*Specimen J* was also lifted out from the b-type hydride grown on the opposite side of the grain boundary, as shown in Figure 8 a. Figure 9 a shows the corresponding tomographic reconstruction, which contains mostly α-Zr, and a small hydride induced by sample preparation (in grey). Three Sn-rich plates are highlighted by an iso-composition surface with a threshold of 2at%. In addition, a planar feature akin to an interface or a grain boundary enriched with oxygen crosses the entire field of view, as highlighted by a set of isocomposition surfaces with a threshold of 2at% O. These isosurfaces also evidence a set of small O-rich clusters. These features are also evidenced in the set of top-views on the 5 nm slices labelled (1)–(3) displayed in Figure 9 b. Conversely to the stacking faults in Figure 8, only two of the planar features are parallel to each other, and another one appears to be located on another set of planes apparently located at approx. 120 degrees from each other. The composition profiles plotted in Figure 9 c were calculated perpendicularly to the SFs, revealing an enrichment in Sn from 2 to 3 at. % in a matrix that appears slightly depleted in Sn, similar to previously discussed. The partial crystallographic information within these APT datasets is insufficient to allow for a full identification of the planes on which the SF are located or if the interface is e.g. a grain boundary.

In Figure 10 a, slice (2) is reproduced and displayed alongside a schematic view showing four distinct regions. A composition profile, Figure 10 b, shows that the composition in Sn and Fe are constant, whereas the O composition in these four regions is different. We extracted yz cross-sections in each of these regions, as displayed in Figure 10 c. In region (I), the O distribution appears homogeneous, with a composition of approx. 0.85 at. %. In (II), within the oxygen-rich interface, the composition reaches 1.7 at. % O, and the O seem homogenously distributed. In (III), the composition in O is only approx. 0.5 at. % , and some of the oxygen is concentrated in oxygen-rich particles that appear mostly aligned. In Region (IV), the average composition is close to only 0.6 at%, with O concentrated within small oxygen-rich particles.

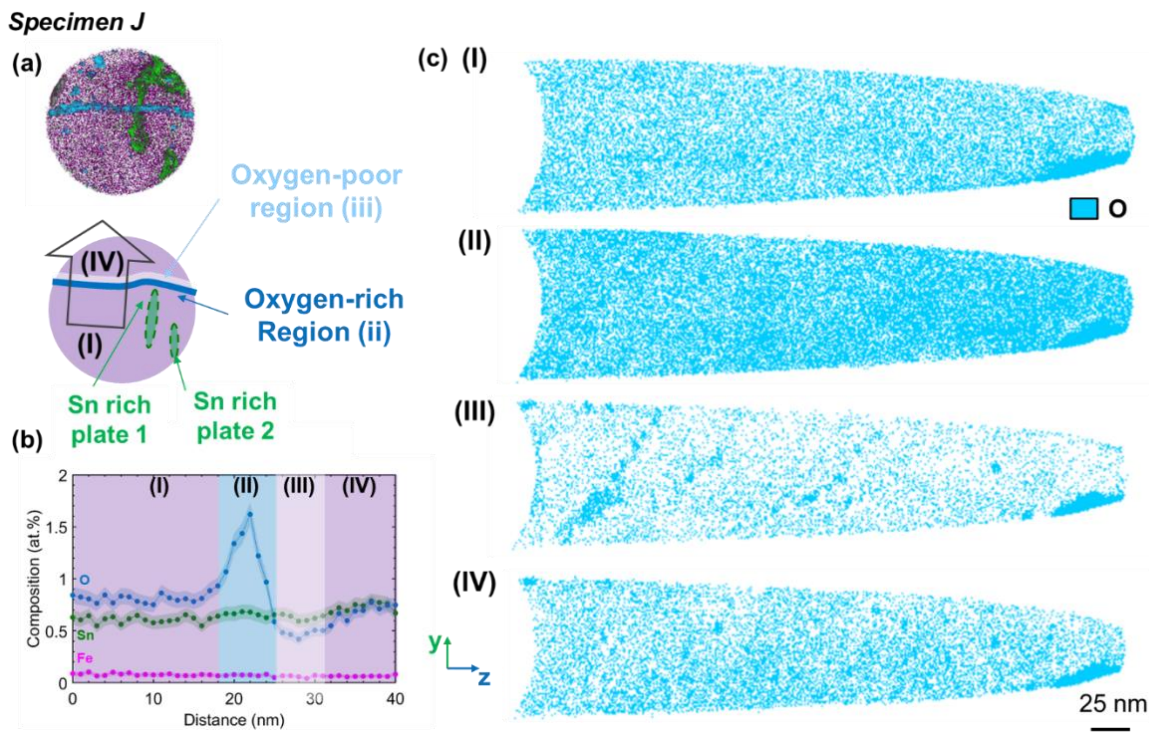

**Fig. 10.** (a) Slice along xy from APT reconstruction of the *specimen J*, in Fig 9 in an **Intergranular hydride (type-b) in H/D charged blocky-α Zircaloy-4** along with a schematic view depicting the microstructure. (b) Composition profiles taken perpendicularly to the oxygen-rich plane (the arrow indicated in (a). (c). 5nm-thick slice along yz from within the APT reconstruction highlighting the O distribution in the 4 zone defined in the composition profile (b).

## 4. Discussion

We have investigated microstructural features from the as-received sample and in the blocky-α microstructure, before and after electrochemical charging with hydrogen/deuterium followed by a low temperature heat treatment at 400 °C for 5 hours followed by furnace cooling at a rate of 0.5 °C/min. Of particular interest, our results include the composition and solute distributions in the bulk of the grains, i.e. the metallic matrix, at different grain boundaries, within various types of hydrides as well as the hydride-matrix interface. In combination, this allows us to discuss in more details the evolution of the solute distribution over the course of the sample's thermal history as well as that caused by the growth of the hydrides. Sn segregates at the hydride/matrix interface. In the case of the b-hydrides, that are grown fast inside the metallic matrix, we report stacking faults formed ahead of the growth front, that are segregated with Sn. The information we provided through this thorough and rather systematic investigation enables to reconsider some important aspects of how the solute distribution affects the alloy's lifetime in service. More detailed discussions on some of these aspect follows, in particular on the possible influence of the segregation to faults on further hydride growth.

### 4.1. Solute H/D distribution

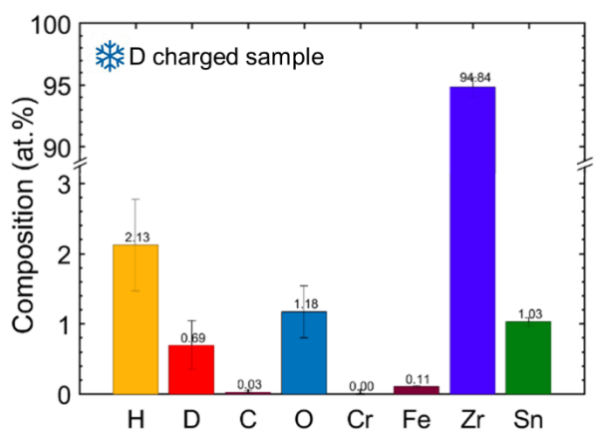

**Fig. 11.** Composition of the matrix after peak decomposition in the *specimen D, cryo-prepared and D-charged*

The distribution of solute hydrogen in the matrix is challenging by APT, even when D-charging is used. Quantitative analysis by APT is extremely challenging in such metallic systems [31], in particular when laser-pulsing mode is used [35] as a significant proportion of the hydrogen to be field evaporated as $H_2^+$ and overlap with the peak of deuterium, hindering quantification. To address this, we analysed a D-charged blocky-α sample, prepared at cryogenic temperature and using cryogenic transfer and we estimated the composition using the deconvolution protocol we introduced in Ref.[31]. The data is reported in Fig. 11. For this analysis, peak decomposition was performed in voxels of $5nm^3$.

In specimen D, the estimated composition in D is around 0.69±0.35 at%. The uncertainty reported here is associated with the standard deviation of the distribution of D within the set of voxels. We note that this D composition may be an over estimation, since over 2 at% of H are detected and it is unclear what the H2/H ratio would be in the conditions of electrostatic field used during this experiment. However, we observe a significant fraction (~33%), of the D comes from the detection of $ZrD^{2+}$ molecular ions.

To verify this, we compared a H-charged and a D-charged sample, both prepared under cryogenic conditions and with comparable Zr charge-state ratios. In these comparisons, we note that that $ZrH_2^{2+}$ is consistently unlikely to appear, and makes up less than 0.3% of the H-containing species. For the D-charged sample, this gives us a lower bound of the amount of D left in the matrix, approx. 0.22 at%. Importantly, this means that even with a slow cooling allowing for extensive precipitation of hydrides, there remains enough D in solution to promote further precipitation over time. Indeed, the solubility at room temperature is 1 weight ppm, i.e. approx. 100 appm so less than 20 times less than what we measured [6].

### 4.2. Distribution of Fe, Sn and O

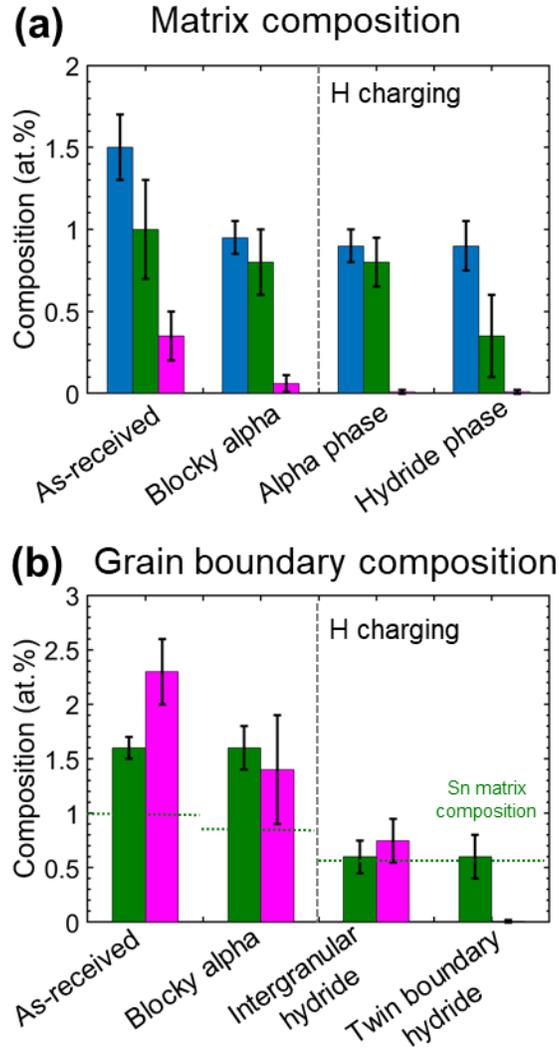

**Fig. 12.** (a) matrix and (b) grain boundary compositions in O, Sn and Fe in the as-received and blocky-α samples before and after H-charging.

In Fig. 12 (a), we report the matrix composition in O, Sn and Fe for the as-received, blocky- α and H/D-charged blocky-α samples. In this last case, we separated the matrix and the hydride. The Sn composition in the α-Zr matrix does not appear to significantly change across the different samples, with a small reduction observed in the hydride, as already reported [30]. The change in the Fe composition in between the as-received and blocky-α microstructures may be explained by the coarsening of Fe-rich secondary-phase particles (SPP) [39,40] – which we did not capture here – during the high-temperature heat treatment, i.e. at 800°C for two weeks. The absence of any significant levels of Cr from our analyses can also be ascribed to Cr being mostly in SPPs already from the as-received state, and again the Cr-rich SPPs were not analysed here. There are remaining Fe-rich clusters within the matrix, indicative of a rather low solubility of Fe in α-Zr. Similar clusters have been previously reported in a similar alloy [23]. These are also found within the hydrides. APT does not have the spatial resolution to reveal whether the clusters have adopted a different crystal structure when in the hydride, but they appear to be sufficiently stable to remain unaffected by the change in the surrounding crystal structure.

We observe a drop in the oxygen content between the as-received and blocky-α states, from approx. 1.5 at% to approx. 1 at%. Precise quantification of O by APT in such systems is a herculean task beyond the scope of the current paper, and these values should only be seen as indicative. Indeed, O appears as a typical impurity in APT, partly due adsorbed species ($O_2$, $H_2O$…) on the surface during specimen transport though air, or simply from the residual gas. The ratio of the charge state of Zr, indicative of the electrostatic field conditions over the course of the APT data acquisition, is sufficiently similar between datasets to allow for comparison. The lower O matrix composition may be related to O segregating to crystalline defects, as seen in Fig.8 for instance, and hence being depleted from the matrix. In the case of the data in Fig. 10, these segregations could be associated to populations of geometrically-necessary dislocations or sub-grain or low-angle grain boundary that remain within the grains following the heat treatment to generate the blocky-α microstructure, as imaged in Tong and Britton [17]. The diffusion and trapping of oxygen to these defects is driven by the minimisation of the system's free energy. The oxygen likely stabilises these crystalline defects and prevent their annihilation at other GBs.

In Fig. 12 (b), we report the composition in Sn and Fe at the imaged grain boundaries in the as-received, blocky- α and H/D-charged blocky-α samples. We do not have access to all the parameters of each analysed grain boundary, i.e. misorientation and grain boundary plane, and these values should be seen as indicative. The data show that Sn and Fe both have a tendency to segregate, except to the twin boundary. Twins are typically low-energy boundaries, and since it likely formed late in the sample's history, there was little time and less thermal activation (and possibly driving force) for solutes to segregate. Fe and Sn segregation to GBs was reported previously in Zircalloy 4 [23]. In *Specimen I* (Fig. 8), O was also seen segregated to a GB between two α-Zr grains at which no Fe or Sn appeared to be segregated. This was the only GB for which these two solutes were no seen segregated apart from the twin in *Specimen G* (Fig.6). This could indicate a specific interaction between O and these solutes at GBs. This could affect the possible pick up of O and hence affect their resistance to corrosion.

### 4.3. Hydride growth

As discussed above and by Wang et al., there are two types of grain boundary hydrides [18], classified based on their growth parallel or normal to the boundary and referred to respectively as a-type and b-type. Note that the hydride studied in our preliminary work, ref. [22], was an a-type hydride, similar to those seen in *Specimen F* for instance. The hydride in *Specimen I* however is of b-type. In both cases, as showed in Fig. 13 (a)-(b) the variation in the H-composition is indicative of the presence of a the intermediate, metastable ζ-hydride phase, as reported in ref. [22]. The behaviour of Sn in all cases agrees with that indicated in ref. [22], it is repelled from the hydride. Fe and Sn were found segregated at GB in α-Zr in both the as-received fine grain and in the annealed blocky-α samples with large grains [17]. Interestingly, looking at Fig.7 and Fig. 12 (b), it seems that we observe systematically less Sn at GBs following the hydride growth. This can be explained if the Sn initially segregated to the GB is pushed away from the GB by the growing hydride. This change in the GB chemistry likely affects its properties, including the oxidation resistance. This contrasts with Fe that is consistently found segregated, including at GB where hydrides have grown. Both a- and b-type hydrides are found to contain the small Fe-rich clusters also found in the matrix, which indicated that they are getting incorporated within the hydride.

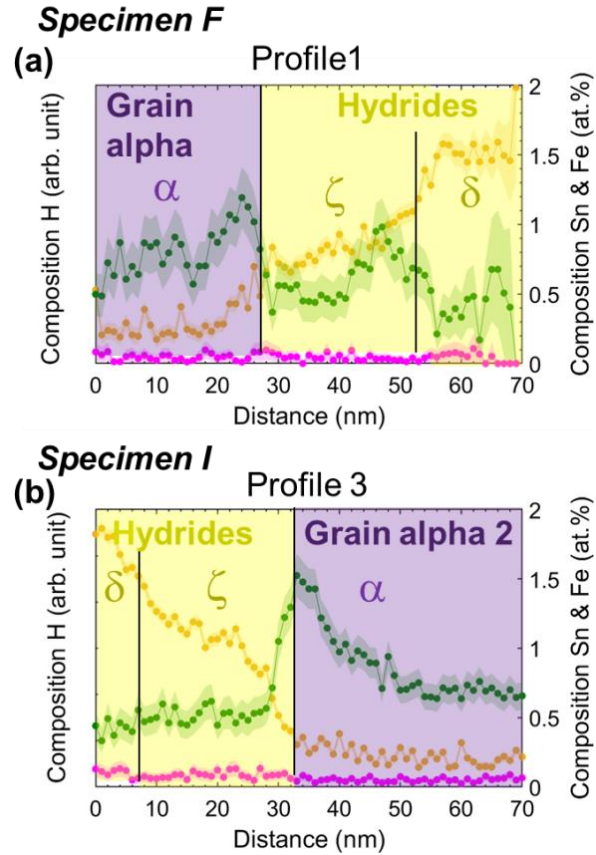

**Fig. 13.** Interfaces between the alpha phase and the hydride.

A striking difference between a- and b- type hydrides is the presence of stacking-faults within the Zr matrix ahead of the growth front in the b-type hydride. The interface of the a-type reported here and in ref. [22] did not exhibit these stacking faults ahead of the growth front, and only shows Sn segregation. Defects created ahead of the hydride growth front have previously been reported. Indeed, the volume expansion associated to the hydride compared to the matrix generates a high local strain around the hydride, which is accommodated in part by creating structural defects at the hydride–matrix interface. Carpenter et al. observed by TEM that dislocation loops in the basal plane {0001} attached to the precipitated hydrides [37][41]. Shinohara et al. [42] found that such dislocation loops formed couples of two neighboring dislocations which moved in parallel. This indicates that the emitted dislocations extend into partial dislocations. Interestingly, the orientation relationship between the matrix and the hydride is the same ($\{0001\}_\alpha||\{111\}_\delta, <11\bar{2}0>_\alpha\ ||\ <110>_\delta$) for both a- and b-type intergranular hydrides [17], indicating that the difference in growth mechanisms for these two types of hydrides does not lie in the crystallography.

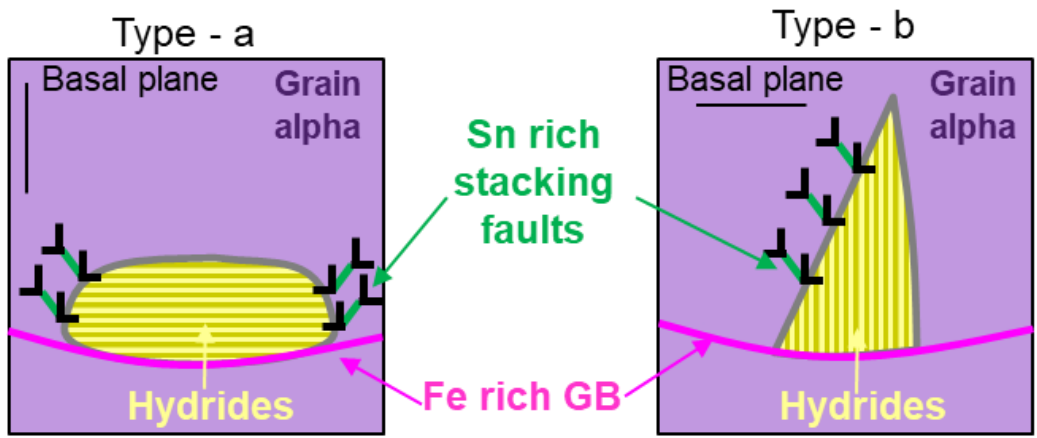

**Fig. 14.** Schematic illustration of Sn rich stacking faults position in a and b type hydride.

Here we reveal segregation of Sn to these stacking faults. Figure 14 is a schematic depiction of the microstructure of both the a- and b-type hydrides. The a-type hydrides grow faster parallel to the grain boundary, and the SFs should appear on the sides that are normal to the GB itself. Since in the material analyzed herein multiple a-type hydrides have grown and formed hydride packets, these stacking faults would then end up either within the hydride or annihilated. Wang et al. previously reported that α-Zr grains within the packets, as seen here in Fig.7 for instance, exhibit a difference in orientation compared to the parent grain, which can be attributed to the strain accumulation, originating, at least in part, from the mismatch between the hydride and the α-Zr matrix. The stacking fault in Fig 6., fully embedded within this a-type hydride packet, indicates as well that the SF can end up integrated within the hydride itself. For b-type hydrides, the growth perpendicular from the GB appears faster, and the SF ahead of the growth front, within the α-Zr matrix, where we observe them.

Alloying had been proposed to lowers the stacking fault energy [37] and by using atomistic simulations, Udagawa et al. showed that Sn lower the stacking fault energy in α-Zr [43].Domain et al.[44] also used atomistic simulations to study the interaction of hydrogen with stacking faults in a Zr–H solid solution. Their results indicate that H could reduce the stacking fault energy and favor the spreading of the dislocation core in the prismatic plane. O, as seen in Fig. 10 could also maybe play a role and help stabilize crystalline defects. Overall, this atomistic calculation work rationalizes our observations of Sn segregating to SFs. Udagawa et al. [43] also discussed how the local change of atomic packing from hcp to fcc at the stacking fault facilitates the nucleation of the hydride. This explains why twin boundaries, with structures that are close to stacking-faults can act as preferential nucleation sites for hydride formation. In addition, the defects inside grains, as imaged in Fig. 10, could also act as the heterogeneous nucleation sites of intragranular hydrides, which are more commonly observed in the blocky- α microstructure than in the fine-grained microstructure as reported by Birch et al. [20]. The smaller number/area of grain boundaries per unit volume in the large grain material also likely plays a role. The formation and stabilization of the stacking faults ahead of the growth front of the b-type hydrides can also affect the hydride growth kinetics since the further formation of a hydride phase in front of the existing hydride is facilitated by the presence of the defect. This could explain why the b-type hydrides grow into the α-Zr matrix farther from the grain boundary, than the a-type hydrides.

## 5. Conclusion

In order to enable significantly better understanding of the hydride growth mechanism in Zircaloy-4, we have performed a thorough analysis of grain boundaries and the bulk of fine-grained and blocky-α microstructures, and after electrochemical hydrogen or deuterium charging followed by heat treatment and slow cooling in the furnace. We discuss issues pertaining to the in-situ formation of additional hydrides associated to the FIB-based specimen preparation, which can be avoided by cryo-FIB, as well as how we approached this issue in how we processed and interpreted the data. We can for instance confidently say that there is D left in the matrix after the slow-cooling process, yet precise quantification of the D-content in the matrix and various precipitated hydrides remains challenging. We reveal that Fe and Sn segregate to most grain boundaries in each of the samples – to the notable exception of the twin boundary. The Fe distribution does not appear to be significantly affected by the growth of the hydride. The Sn is however repelled from the growing hydride, as reported previously. For both the a- and b-type hydrides, a layer of intermediate composition forms. For the b-type hydride, which grow near perpendicularly to the initial grain boundary, stacking faults are imaged ahead of the growth front, and found to capture some of the Sn. These stacking faults can then assist further with the growth. O is also found at defects in the bulk of the grain and at a grain boundary where Fe and Sn are not detected. Overall, our results shed light onto the role and behavior of the some of the typical alloying additions in these systems. Sn could well play a major role in slowing down the growth of hydrides by segregating to the interface. Yet by stabilizing stacking faults forming ahead of the growth front of some of the hydrides, it may also accelerate the growth as well. These complex processes likely need some more high-resolution, in-situ work to be fully understood, alongside some atomistic simulations.

## Acknowledgements

Uwe Tezins, Chris Broß, and Andreas Sturm for their support to the FIB and APT facilities at MPIE. The authors are grateful to Agnieszka Szczepaniak for her help and support with the preparation of the specimens, esp. her heroic efforts with the cryo-PFIB – she asked not be a co-author on the article. IM, LTS and BG are grateful for the Max-Planck Society and the BMBF for the funding of the Laplace and the UGSLIT projects respectively, for both instrumentation and personnel. BG and LTS are grateful for financial support from the ERC-CoG-SHINE-771602. TBB thanks the Royal Academy of Engineering for support of his Research Fellowship. TBB and SW acknowledge support from HexMat (EP/K034332/1) and MIDAS (EPSRC EP/SO1720X) programme grants.